\documentclass[twocolumn]{aastex61} 

\usepackage{aas_macros}          
\usepackage{mathtools}           
\usepackage{cases}               

\defcitealias{2011ApJ...738...71S}{SP11}
\defcitealias{2013ApJ...776..134P}{PBS13}
\defcitealias{2007ApJ...657L.113L}{Langton \& Laughlin 2007}
\defcitealias{2010GeoRL..3718811S}{Showman \& Polvani 2010}

\usepackage{tikz}

\usepackage{printlen}

\newcommand\beq{\begin{equation}}   
\newcommand\eeq{\end{equation}}  
\def\bal#1\eal{\begin{align}#1\end{align}}  
\newcommand\bse{\begin{subequations}}    
\newcommand\ese{\end{subequations}}       

\def\bml#1\eml{\begin{multline}#1\end{multline}}  
\def\bspl#1\espl{\begin{split}#1\end{split}}  

\newcommand\bga{\begin{gather}}    
\newcommand\ega{\end{gather}}      

\usepackage{xcolor}
\definecolor{mygray}{gray}{0}

\newcommand\mygray[1]{{\color{mygray}#1}}
\newcommand\eqnref[1]{Equation (\ref{#1})}    
\newcommand\figref[1]{Figure \ref{#1}}      
     
\newcommand\sectref[1]{Section \ref{#1}}     

\newcommand\ReynMag{\mathrm{Rm}}

\newcommand\kelvin{\, \mathrm{K}}	
	
\newcommand\persecond{\, \mathrm{s}^{-1} }	
\newcommand\Eday{\,\, \mathrm{Earth}\, \mathrm{day} }	
	
\newcommand\metre{\, \mathrm{m}}	
\newcommand\kilometre{\, \mathrm{km}}

\newcommand\gauss{\, \mathrm{G}}	
\newcommand\kilogauss{\, \mathrm{kG}}	
\newcommand\onebar{\, 1 \, \mathrm{bar}}	

\usepackage{gensymb}     
\newcommand\pa{\partial}

\newcommand\vect[1]{\mathbf #1}                     
\newcommand\uvect[1]{\widehat{\mathbf #1}}  		
\newcommand\di[1]{\mathrm{#1}}					 	
\newcommand\dd{\mathrm{d}}						 	

\newcommand\deriv[2]{ \frac{\partial #1}{\partial #2} }		
\newcommand\lagderiv[1]{\frac{\mathrm{d}#1}{\mathrm{d}t}}

\newcommand{\e}{\mathrm{e}} 				

\newcommand\tdrag{\tau_\di{drag} }	
\newcommand\trad{\tau_\di{rad} }

\newcommand\tTransfer{\tau_\di{transf}}

\received{December 3, 2018}
\revised{February 8, 2019}
\accepted{February 10, 2019}
\published{February 21, 2019}
\submitjournal{ApJL}

\shorttitle{Shallow water magnetohydrodynamics for westward hotspots on hot Jupiters}
\shortauthors{Hindle et al.}

\begin{document}

\title{Shallow-water magnetohydrodynamics for westward hotspots on hot Jupiters}

\correspondingauthor{Alexander Hindle}
\email{a.hindle@newcastle.ac.uk}

\author[0000-0001-6972-2093]{A. W. Hindle} 
\affil{School of Mathematics, Statistics and Physics, Newcastle University, Newcastle upon Tyne, NE1 7RU, UK}

\author[0000-0002-4691-6757]{P. J. Bushby}
\affil{School of Mathematics, Statistics and Physics, Newcastle University, Newcastle upon Tyne, NE1 7RU, UK}

\author[0000-0002-2306-1362]{T. M. Rogers}
\affil{School of Mathematics, Statistics and Physics, Newcastle University, Newcastle upon Tyne, NE1 7RU, UK}
\affil{Planetary Science Institute, Tucson, AZ 85721, USA }

\begin{abstract}    
Westward winds have now been inferred for two hot Jupiters (HJs): HAT-P-7b and CoRoT-2b. Such observations could be the result of a number of physical phenomena such as cloud asymmetries, asynchronous rotation, or magnetic fields. For the hotter HJs magnetic fields are an obvious candidate, though the actual mechanism remains poorly understood. Here we show that a strong toroidal magnetic field causes the planetary-scale equatorial magneto-Kelvin wave to structurally shear as it travels, resulting in westward tilting eddies, which drive a reversal of the equatorial winds from their eastward hydrodynamic counterparts. Using our simplified model we estimate that the equatorial winds of HAT-P-7b would reverse for a planetary dipole field strength $B_{\di{dip},\text{HAT-P-7b}}  \gtrsim 6 \gauss $, a result that is consistent with three-dimensional magnetohydrodynamic simulations and lies below typical surface dipole estimates of inflated HJs. The same analysis suggests the minimum dipole field strength required to reverse the winds of CoRoT-2b is $B_{\di{dip},\text{CoRoT-2b}} \gtrsim 3 \kilogauss $, which considerably exceeds estimates of the maximum surface dipole strength for HJs. We hence conclude that our magnetic wave-driven mechanism provides an explanation for wind reversals on HAT-P-7b; however, other physical phenomena provide more plausible explanations for wind reversals on CoRoT-2b.    
\end{abstract}


\section{Introduction} \label{sec:intro}
Observations of hot Jupiters (HJs) generally measure a peak brightness offset eastward of the substellar point \citep{2009ApJ...690..822K, 2016ApJ...823..122W}. Similarly, equatorial superrotation is an archetypal feature of hydrodynamic models of tidally locked, strongly irradiated, short-period planets \citep{2002A&A...385..166S,2005ApJ...629L..45C,2007ApJ...657L.113L,2008ApJ...673..513D}.  Furthermore, \cite[][hereafter \citetalias{2011ApJ...738...71S}] {2011ApJ...738...71S} showed that such systems will always produce eastward equatorial jets, which are driven by interactions between the mean flow and the system's linear equatorial shallow-water hydrodynamic (SWHD) waves.  However, recent continuous Kepler measurements of HAT-P-7b and thermal phase observations of CoRoT-2b made by the Spitzer Space Telescope found westward-venturing peak brightness and hotspot offsets \citep{2016NatAs...1E...4A,2018NatAs...2..220D}. These observations suggest the existence of a mechanism that can also drive westward equatorial winds.
 
 Based on their magnetohydrodynamic (MHD) simulations, \cite{2014ApJ...794..132R} predicted that westward wind variations would occur as the result of strong coupling between a planet's flow and magnetic field. Furthermore, \cite{2017NatAs...1E.131R} highlighted that, assuming wind reversals are magnetically driven, observations of westward hotspot offsets lead to a direct  constraint on the magnetic field strengths of a given HJ. While \cite{2017NatAs...1E.131R} demonstrated that westward flows developed in the strong field case, the actual mechanism for wind reversals remained unknown.

Here we demonstrate that a shallow-water wave-driven mechanism can explain the wind reversals. Firstly, we demonstrate that a shallow-water magnetohydrodynamic (SWMHD) model can reproduce both eastward hotspot offsets in hydrodynamic cases and westward hotspot in the presence of a strong toroidal magnetic field, suggesting that magnetically driven wind reversal is a shallow phenomenon. We then highlight magnetic modifications to equatorial SWMHD waves and present a wave-driven reversal mechanism, which is consistent with the hydrodynamical theory of \citetalias{2011ApJ...738...71S}. We conclude by discussing the possible consequences of these concepts for HAT-P-7b and CoRoT-2b.

\section{Reduced-gravity SWMHD model}
\mygray{We adapt the SWMHD model of \cite{2000ApJ...544L..79G} and use a reduced-gravity SWMHD model. This is the MHD analog of the reduced-gravity SWHD models used to study HJs in hydrodynamic systems (e.g.,~\citetalias{2007ApJ...657L.113L};~\citetalias{2010GeoRL..3718811S};~\citetalias{2011ApJ...738...71S}).

   \begin{figure}    
   \includegraphics[scale=1]{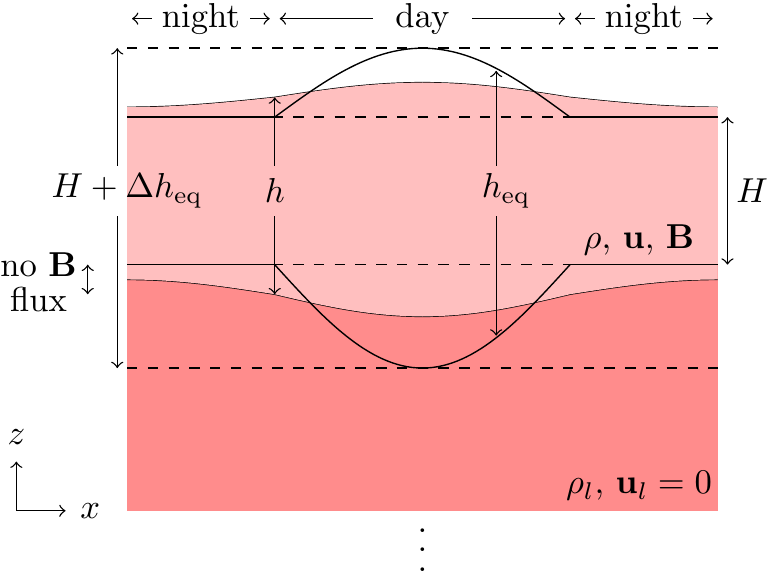}
    \caption{ The reduced-gravity SWMHD model has two layers: an active layer sits upon an infinitely deep inactive fluid layer, where both layers have constant densities ($\rho$ and $\rho_l$, respectively). No magnetic flux is permitted across the interface. The layer thickness, $h$, is relaxed toward the imposed radiative equilibrium thickness profile, $h_\di{eq}$, over a radiative timescale, $\trad$.}
    \label{fig:model}
\end{figure} 

The reduced-gravity SWMHD model, as illustrated in \figref{fig:model}, has two constant density layers: an upper, meteorologically active layer and an infinitely-deep, quiescent lower layer. In the absence of forcing the active layer has a thickness $H$, which is physically analogous to the pressure scale height. 

In the limit $H/L \ll 1$, where $L$ is some typical horizontal length scale, vertical acceleration becomes vanishingly small and the system lies in magneto-hydrostatic balance: gravitational acceleration balances the total (gas plus magnetic) vertical pressure gradient, and the horizontal velocity and magnetic fields become independent of the vertical coordinate, $z$. Consequently, the MHD equations can be integrated over $z$ (while requiring that interfaces between vertical layers are material surfaces, with no magnetic flux across them) to give the reduced-gravity SWMHD equations. For a local Cartesian system in the equatorial beta-plane approximation, the evolution of the active layer of the reduced-gravity SWMHD model is governed by the equations
 \begin{align}
 \begin{split}
    \lagderiv{ \vect{u} } + \beta y (\uvect{z} \times  \vect{u})  & = - g \nabla h + ( \vect{B} \cdot \nabla) \vect{B} 
    \\  & \quad + \vect{R} - \frac{ \vect{u} }{\tdrag} + \vect{D}_\nu, 
    \label{eqn:mom}
 \end{split}
\\
    \deriv{h}{t} + \nabla \cdot (h \vect{u} ) & = \frac{ h_\di{eq} - h }{\tau_\di{rad}} \equiv Q,   \label{eqn:cont}
\\
\lagderiv{ A }  &=  D_\eta, \label{eqn:ind}
\end{align}
where $h(x,y,t)$, $\vect{u}(x,y,t)\equiv(u,v)$, and $\vect{B}(x,y,t)\equiv(B_x, B_y)$ denote the active layer thickness, the horizontal active layer velocity field, and the horizontal active layer magnetic field (in units of velocity), respectively. The horizontal gradient and Lagrangian time derivative operators are defined by $\nabla \equiv (\pa_x, \pa_y)$ and $\di{d}/ \di{dt} \equiv \pa/ \pa t + \vect{u} \cdot \nabla$, respectively. 

The system is defined in terms of a magnetic flux function, $A(x,y,t)$, which satisfies $h \vect{B} = \nabla \times  A \uvect{z}$, thus guaranteeing that the SWMHD divergence-free condition, $\nabla \cdot (h \vect{B}) =0$, remains satisfied everywhere for all time. We take the system's origin $(x,y)=(0,0)$ to be the modeled planet's substellar point, therefore our system is compared to spherical geometries with the approximate coordinate transforms $\phi \approx x/R$ and $\theta \approx y/R$ (where $\phi$ and $\theta$ denote the azimuthal and latitudinal coordinates, and $R$ denotes the planetary radius). The reduced gravitational acceleration is denoted by the constant $g$, which is defined as in \cite[][hereafter \citetalias{2013ApJ...776..134P}]{2013ApJ...776..134P}  rather than \cite{2006aofd.book.....V}, and the latitudinal variation of the Coriolis parameter at the equator is given by } $\beta \equiv \dd f / \dd y |_{y=0} = 2 \Omega / R$, for planetary rotation frequency $\Omega$.

In numerical simulations we include explicit viscous diffusion \citep{doi:10.1093/qjmam/hbu004} 
\beq
\vect{D}_\nu = h^{-1} \nabla \cdot \left[ \nu h \left(\nabla \vect{u} + (\nabla \vect{u})^T \right) \right],
\label{eqn:viscdiff}
\eeq
where $\nu$ is the kinematic viscosity. \mygray{Furthermore, we treat the induction equation with the explicit magnetic diffusion (A.~D.~Gilbert et al.~2019, in preparation)
\beq
D_\eta =  \eta (\nabla^2 A-  h^{-1} \nabla h \cdot \nabla A), \label{eqn:magdiff}
\eeq
where $\eta$ is the magnetic diffusivity. 

The prescribed Newtonian cooling term, $Q$, relaxes the system toward the imposed radiative equilibrium profile, $h_\di{eq}$, over a radiative timescale, $\trad$, by transferring mass upward from the infinitely deep inactive layer to the active layer in ``heating'' regions and vice versa in ``cooling'' regions. 

The vertical mass transport, $\vect{R}$, represents the effect of Newtonian cooling on the momentum equations.  In cooling regions ($Q < 0$) mass is transported downward into the infinitely deep inactive layer, and the specific momentum of both layers is conserved without any horizontal acceleration. Conversely, in regions of heating ($Q > 0$) mass with no horizontal velocity is transported upward into the active layer, causing the horizontal deceleration of the active layer. In heating regions it is required that Newtonian cooling has no effect on the temporal evolution of the specific momentum, $\pa (h \vect{u}) / \pa t$, hence, from Equations \eqref{eqn:mom} and \eqref{eqn:cont}, $h \vect{R} + \vect{u} Q$ must sum to zero, giving
\begin{equation}
    \vect{R}=
  \begin{cases}
        0  &  \di{for}\,\, Q < 0 \\
        -  \frac{ \vect{u} Q}{h}  & \di{for}\,\, Q \geq 0 ,
  \end{cases}
\end{equation}
which has previously been used in comparable SWHD models (e.g.,~\citetalias{2010GeoRL..3718811S};~\citetalias{2011ApJ...738...71S};~\citetalias{2013ApJ...776..134P}). We also include simple Rayleigh drag in \eqnref{eqn:mom} for direct comparison with these SWHD models.}

\section{Numerical treatment and solutions} \label{sect:numerical:treatment:solutions}

\mygray{We evolve the system by solving Equations \eqref{eqn:mom}--\eqref{eqn:ind} on the domain $-\pi < x/R < \pi$, $-\pi/2 < y/R < \pi/2$, from a flat rest state ($h=H$ and $\vect{u}=\vect{0}$ everywhere) for SWHD solutions, then impose a background magnetic field ($A=A_0$) once a hydrodynamic steady state is achieved for SWMHD solutions.

We apply periodic boundary conditions on $\vect{u}$, $h$ and $A$ in the $x$ direction and require $v = 0$,  $\pa u / \pa y = 0$, $\pa A / \pa x = 0$ ($h$ is chosen to conserve mass) at $y$ boundary points. The equations are solved on a $256 \times 255$ grid in $x$ and $y$, with spatial derivatives taken pseudo-spectrally in $x$ and using fourth-order finite difference schemes in $y$. We integrate the system forward in time using an adaptive third-order Adam--Bashforth scheme \citep{2003ApJ...588.1183C}.

The system is driven by relaxing $h$ toward the prescribed radiative equilibrium layer thickness profile 
  \beq
    h_\di{eq} = 
 \begin{cases}
    H + \Delta h_\di{eq} \cos \left( \frac{x}{R} \right) \cos \left( \frac{y}{R} \right)    & \di{dayside} \\
    H  & \di{nightside},\\
  \end{cases} \label{eqn:heqprof}
\eeq
where $H$ is the nightside equilibrium thickness and $\Delta h_\di{eq}$ is the difference in $h_\di{eq}$ between the nightside and the substellar point. This profile is the Cartesian analog of the spherical forcing prescription used in comparable hydrodynamic models (e.g.,~\citetalias{2007ApJ...657L.113L};~\citetalias{2010GeoRL..3718811S};~\citetalias{2011ApJ...738...71S}). 

HJs have weakly ionized photospheres. Consequently, strong zonal flows crossing the assumed deep-seated planetary dipolar magnetic field are believed to induce atmospheric toroidal fields. \cite{2012ApJ...745..138M} showed that the strengths of the dipolar field, $B_\di{dip}$, and the toroidal field, $B_\phi$, can be approximately related by the scaling law $B_\phi \sim \ReynMag B_\di{dip}$, where the magnetic Reynolds number ($\ReynMag$) is temperature dependent and exceeds unity for hotter HJs} ($T_\di{eq} \gtrsim 1300 \kelvin $). Hence, in such systems the toroidal field is expected to dominate the dipolar field in equatorial regions.

\begin{figure*}
    \plotone{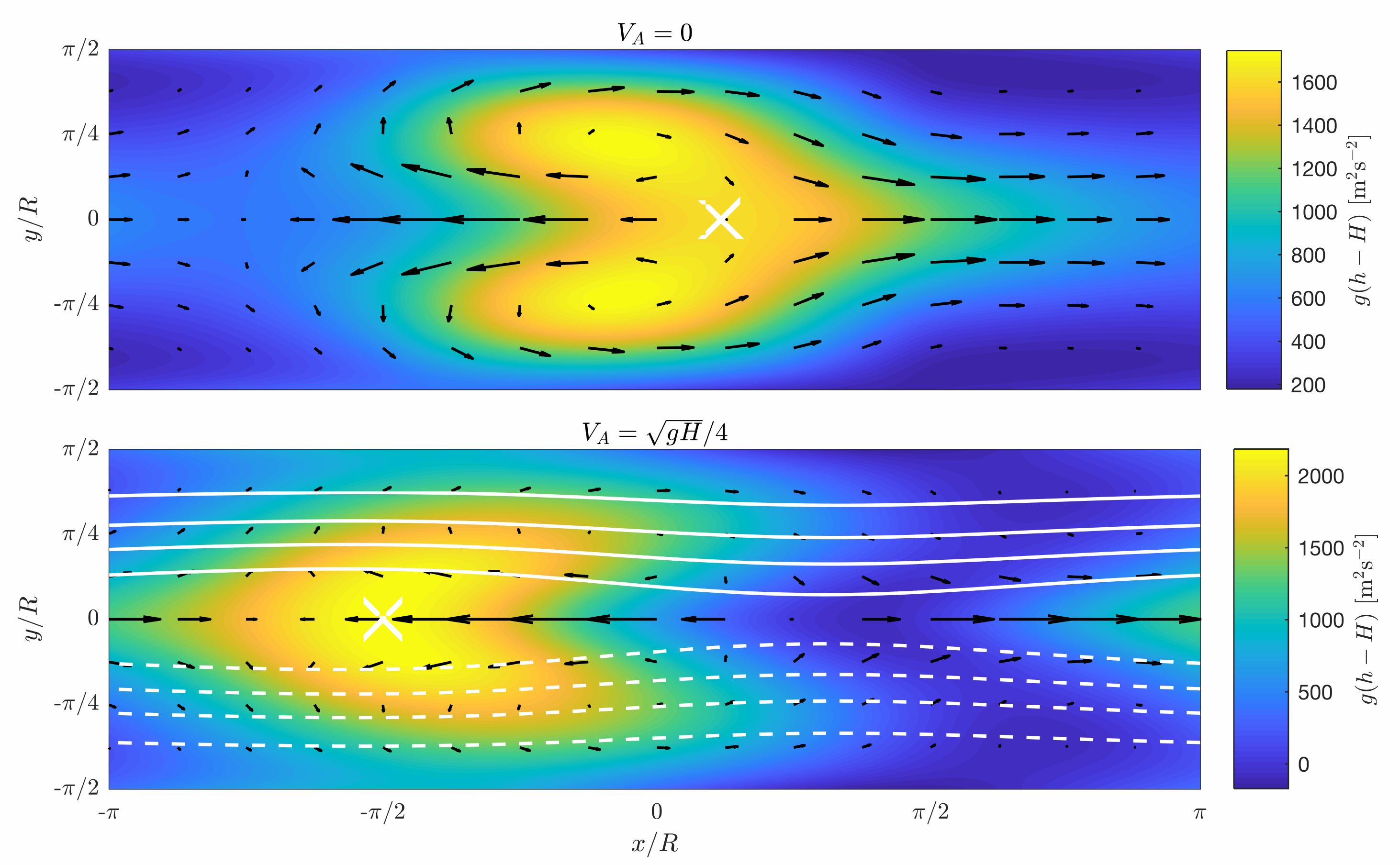} 
\caption{Contours of $g(h-H)$ are plotted for (quasi-)steady solutions, with a forcing amplitude of $\Delta h_\di{eq} / H = 0.001$ (linear regime), and the radiative/drag timescales $\trad = \tdrag = 1 \Eday$. Wind velocity vectors are overplotted as black arrows, lines of constant horizontal magnetic flux ($A$) are overplotted as white lines (with solid/dashed lines representing positive/negative magnetic field values), and hotspots (maxima of $h$ on the equatorial line) are marked by  {white crosses}. The system origin lies at the substellar point and velocity vectors are independently normalized for each subplot.}
   \label{fig:lineargeopotential}
\end{figure*}

Numerically, we implement an equatorially antisymmetric azimuthal background magnetic field though a background flux function, which we impose initially and allow to evolve. The imposed background flux function takes the form
\beq
A_0(y) = - \e^{1/2}H V_\di{A} L_m \e^{- y^2/2 L_m^2},
\eeq
where the background Alfv\'en speed, $V_\di{A}$, determines the background magnetic field strength of the system. We set the latitudinal decay length of the magnetic field to $L_m =L_\di{eq}/2$, where $L_\di{eq} \equiv (\sqrt{gH}/\beta)^{1/2}$ is the equatorial Rossby deformation radius.

Following (\citetalias{2013ApJ...776..134P}), we choose parameters to match those typical of HD~189733b where possible. This HJ has a planetary radius  $R = 8.2 \times 10^7 \metre$, a planetary rotation rate, $\Omega =  3.2 \times 10^{-5} \persecond$, and gravity waves with a speed of $\sqrt{gH} = 2 \kilometre \persecond$ (\citetalias{2013ApJ...776..134P}). The viscous and magnetic diffusivities are assigned the constant values of $\nu = 10^8 \metre^2 \persecond$ and $\eta = 3 \times 10^7 \metre^2 \persecond$, respectively. These are typical values in the radiative zones of HJs but, in reality, the day--night temperature variations on HJs cause longitudinal variations in diffusion coefficients, which can be orders of magnitude for $\eta$. We fix the atmospheric pressure scale height $H=400 \kilometre$ (\citetalias{2013ApJ...776..134P})  and vary the background magnetic field strength via the free parameter $V_\di{A}$, presenting solutions in the weakly forced ($\Delta h_\di{eq} / H=0.001$)  and therefore approximately linear regime, with radiative/drag timescales corresponding to moderately efficient energy redistribution ($\trad = \tdrag = 1 \Eday$; \citetalias{2013ApJ...776..134P}).

After an initial transient period, SWHD solutions reach steady state. For SWMHD systems, the magnetic diffusion timescale is relatively large compared to the system's dynamical timescale ($\tau_\di{dyn}/\tau_\eta  \sim 0.08$) and a {dynamically relevant} quasi-steady state emerges, before diffusion causes the magnetic field to decay. We present numerical SWHD and SWMHD solutions in these steady and quasi-steady states, respectively, and plot $g(h-H)$, the geopotential above the nightside equilibrium reference state, in \figref{fig:lineargeopotential} for  $V_\di{A} = 0$ and $V_\di{A} =\sqrt{gH}/4$ (top/bottom panels, respectively). Energy (heat) redistribution is traced via the geopotential, with high geopotential regions analogous to high temperature regions (\citetalias{2013ApJ...776..134P}). 

Strikingly, the quasi-steady solution for $V_\di{A}=\sqrt{gH}/4$ (lower panel of \figref{fig:lineargeopotential}) exhibits a westward hotspot offset (marked by a white cross). This is in stark contrast to SWHD systems (and SWMHD systems with $V_\di{A} \ll \sqrt{gH}/4$), which always have an eastward hotspot offset.

Solutions in this ``strong field limit'' have larger geopotential gradients, caused by the role of magnetic tension (geopotential gradients increase  {sharply} as $V_\di{A}$ is raised beyond $V_\di{A} =\sqrt{gH}/4$), and the shape of the geopotential contours undergoes a phase transition as the magnetic field is increased: the eastward-pointing chevron-shaped contours, in the zero or weak field regime, transition into the westward-pointing chevron-shaped contours in the strong field limit. Because \citetalias{2011ApJ...738...71S} showed the eastward-pointing chevron-shaped flow patterns to play a major role in the formation of eastward zonal jets, this latter point is of particular interest concerning wind reversals.

\section{The magnetic modification of equatorial shallow-water waves}
In a hydrodynamic study, \citetalias{2011ApJ...738...71S} showed that SWHD systems will always produce eastward equatorial jets that are driven by interactions between the mean flow and the linear equatorial SWHD waves. The dominant standing, planetary-scale equatorial waves induced by day--night thermal forcing are the $n=1$ Rossby and Kelvin waves. The superposition of these modes causes the emergence of eastward-pointing chevron-shaped (geopotential and velocity) phase tilts (e.g.,~\figref{fig:lineargeopotential}, top panel). These cause eddies to pump eastward momentum from high latitudes to the equator, driving an eastward equatorial jet.

The question addressed in this section is how this process is modified in the presence of a magnetic field. We show that magnetism can cause the superposition of planetary-scale, free equatorial SWMHD waves to acquire phase tilts that are opposite in direction to their SWHD counterparts, then link this to a reversal mechanism.

We linearize Equations \eqref{eqn:mom}--\eqref{eqn:ind}, in the absence of forcing, drag, and diffusion, about the background state $\{u_0,v_0,h_0,B_{x,0},B_{y,0}\}=\{0,0,H,B_0(y),0\}$, where $H$ is constant and $B_0 = V_\di{A} y /R$. This system has previously been solved by \cite{2018ApJ...856...32Z}, who studied it in terms of the solar tachocline, and we repeat this analysis for the HJ parameter space (see \sectref{sect:numerical:treatment:solutions} for parameter choices), highlighting important features concerning HJs. 

Perturbations to the background state are separated using the plane wave ansatz, $\{\vect{u}_1,h_1,\vect{B}_1\}=\{\hat{\vect{u}}(y),\hat{h}(y), \hat{\vect{B}}(y)\} \e^{i(kx-\omega t)}$. The resulting ordinary differential equation, (with terms up to $y^2$ only\footnote{The implied assumptions, that $|V_\di{A}^2 k^2 y^2/\omega^2 R^2|\ll1$ and $|V_\di{A}^2 k^2 y^2/(\omega^2-gHk^2) R^2 |\ll1$, remain valid for the discussed solutions.}) is then solved using the boundary conditions, $|v| \rightarrow 0$ as $|y| \rightarrow \infty$, yielding bounded solutions of the form \citep[][]{2018ApJ...856...32Z}
\beq
 \hat{v}_n(y) = H_n(\sqrt{\mu} y)\e^{-(\mu + d)y^2/2}, \label{eqn:eigenfunction}
\eeq
where $d+\mu > 0$, $H_n(\xi)$ is the Hermite polynomial of order $n$, for $n=0,1,2,3, \dots \;$, and

\begin{align}
    \begin{split}
        \mu & = \left( \frac{V_\di{A}^2 k^2}{R^2gH} + \frac{\beta^2}{gH}  + \frac{2 k^3 \beta V_\di{A}^2 }{R^2 \omega (\omega^2- gH k^2)}    \right.  \\
        & \left.  \quad + \frac{ k^3 \beta V_\di{A}^2}{R^2 \omega^3} + d^2 \right)^{1/2}, 
    \end{split} \\
    d & = \frac{gH V_\di{A}^2 k^4}{R^2 \omega^2 (\omega^2- gH k^2) }. \label{eqn:adjustment:fact}
\end{align}
In Equations \eqref{eqn:eigenfunction}--\eqref{eqn:adjustment:fact} the azimuthal wavenumber, $k$, and oscillation frequency, $\omega$, are linked by the dispersion relation
  \beq
 \frac{\omega^2}{gH}-k^2 - \frac{k\beta}{\omega} - \frac{ gH V_\di{A}^2 k^4}{R^2 \omega^2 (\omega^2- gH k^2)} = (2n+1)\mu, \label{eqn:disp}
\eeq
for $n=0,1,2,3, \dots \;$. 

 We solve \eqnref{eqn:disp} using numerical root finding techniques and find that, as in hydrodynamic theory \citep[e.g.,][]{196625}, there are two bounded $n=0$ solutions and three bounded $n \geq 1$ solutions. Completeness is obtained by replacing the missing/third $n = 0$ solution with a magneto-Kelvin solution, which has the characteristic $v = 0$ everywhere and is often called the $n=-1$ mode \citep[e.g.,][]{196625,2018ApJ...856...32Z}. Setting $v = 0$ everywhere and seeking non-trivial solutions to the linearized versions of Equations \eqref{eqn:mom}--\eqref{eqn:ind}, one finds (for $V_\di{A} < (\beta R^2 \sqrt{gH})^{1/2} $) the single bounded magneto-Kelvin solution
\begin{align}
 &\omega(y)  = k \left(gH + \frac{V_\di{A}^2 y^2}{R^2} \right)^{1/2} , \\
 &\hat{u} \propto \exp \left\{  \left( \frac{V_\di{A}^2}{R^2 gH} - \frac{\beta}{\sqrt{gH}} \right)\frac{y^2}{2}  \right\},
\end{align}
where the exponential profile's argument is approximated with accuracy $O( {V_\di{A}^4 y^4}/{R^4 (gH)^2} )$ . 

\begin{figure*}
    \plotone{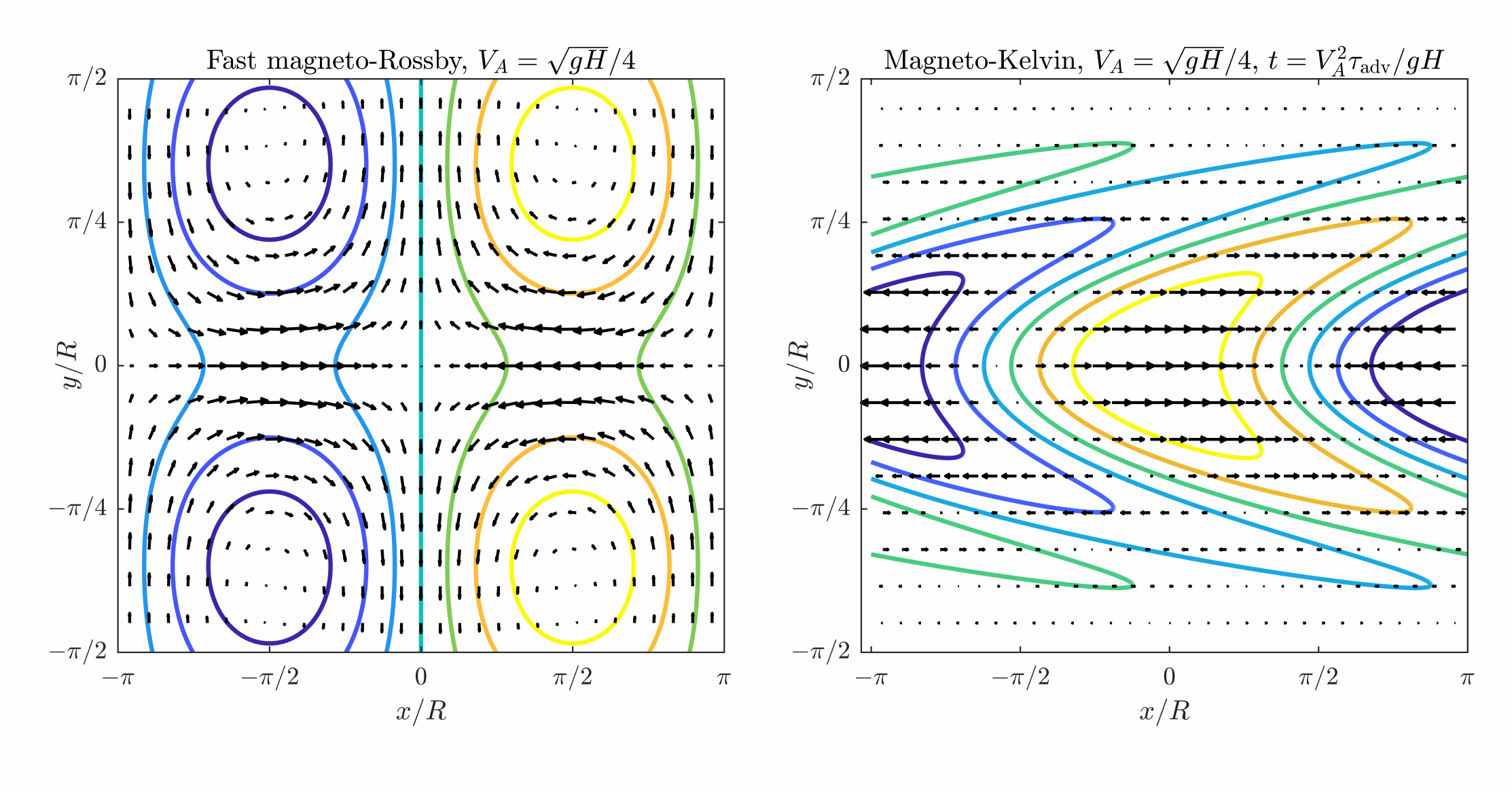}
\caption{Geopotential perturbations contours, $gh_1$, and vectors of velocity perturbations, $\vect{u}_1$, for two different mode types at $k=1/R$. The $n=1$ (fast) magneto-Rossby mode is plotted (left panel) for $V_\di{A}=\sqrt{gH}/4$. The magneto-Kelvin mode is plotted (right panel) for $V_\di{A}=\sqrt{gH}/4$ at $t=\tTransfer \equiv V_\di{A}^2 \tau_\di{adv}/gH$. Plots are made for the parameter choices discussed in \sectref{sect:numerical:treatment:solutions}. } 
  \label{fig:disp_cont}
\end{figure*}

Due to the forms of $h_\di{eq}$, the linearized continuity equation and the eigenfunctions' derivatives \citep[e.g.,][]{1965hmfw.book.....A}, the $n=\pm1$ modes are always expected to play an important role in systems with  $L_\di{eq} \sim R$ (as found on HJs). 
In \figref{fig:disp_cont} we plot geopotential contours of the $n=1$ (fast) magneto-Rossby mode (left panel) and the magneto-Kelvin mode (right panel), for $k=1/R$ (the forcing wavenumber)\footnote{The east/west magneto-inertial gravity waves remain similar to their SWHD forms, which are known to provide an insignificant contribution to linear solutions \citep{196625}.
}.

The structure of the (fast) magneto-Rossby modes do not significantly vary from their hydrodynamic variants \citep[see][]{196625}, although their deformation length does increase with increasing $V_\di{A}$. The magneto-Kelvin mode is the most significantly magnetically modified free wave solution as it acquires a latitudinally dependent contribution to its dispersion relation, which causes the wave to structurally shear as it propagates eastward. We estimate the degree of structural shear in linear quasi-steady solutions by plotting the free magneto-Kelvin wave at $\tTransfer \equiv V_\di{A}^2 \tau_\di{adv} / gH$, the timescale for the wave to transfer a local thickness perturbation, $ h_1$, to surrounding regions in the strong field limit\footnote{We estimate $\tTransfer$ by considering approximate scalings of terms in the continuity equation (${h_1}/ \tTransfer  \sim H U / L$) and momentum equation for a rotationless non-diffusive SWMHD model.  For SWHD models and moderately magnetic SWMHD models, $U/\tTransfer  \sim g h_1/L$, hence $\tTransfer = L_\di{eq}/ \sqrt{gH}$ (\citetalias{2013ApJ...776..134P}). We find numerically that for strong magnetic fields the pressure gradient and Lorentz force approximately balance, yielding $g h_1 / L \sim V_\di{A}^2 / L$, hence $\tTransfer= V_\di{A}^2 \tau_\di{adv} / gH$, where $\tau_\di{adv}\equiv L_\di{eq}/U$ is the hydrodynamic advection timescale defined in \citetalias{2013ApJ...776..134P}.}.

We find that the structural deformation of the magneto-Kelvin wave becomes qualitatively significant at $V_\di{A} \sim \sqrt{gH}/4$. This transition in nature is consistent with the numerical solutions discussed in \sectref{sect:numerical:treatment:solutions}, which also transitions in nature, obtaining a westward-chevron phase shift at $V_\di{A} \sim \sqrt{gH}/4$. 

We hence propose the following adjustment to the mechanism of \citetalias{2011ApJ...738...71S} to account for magnetism: the hydrodynamic mechanism remains valid for low to moderate toroidal field strengths, however, when the toroidal field strength becomes large enough to deform the magneto-Kelvin wave's structure, the resultant superposition of magneto-Kelvin and $n=1$ magneto-Rossby standing waves has a structural form resembling a westward-pointing chevron (such as the one seen in \figref{fig:lineargeopotential}). This change in the wave structure would reverse the sign of the convergence of the meridional flux of zonal eddy momentum. Hence, the structural phase tilts caused by the waves would pump eastward momentum from the equator to higher latitudes and, provided this pumping remains the dominant zonal acceleration process, this would drive a westward equatorial jet.

We comment that this assumption cannot be guaranteed without consideration of the forced linear solutions, which we omit from this Letter, but will investigate in a future paper. 

\section{Discussion}
We have demonstrated that magnetically modified waves lead to westward directed winds in a SWMHD model. We found that the SWMHD model that we presented can capture the physics of magnetically induced wind reversals, which have only previously been studied via full three-dimensional MHD simulations  \citep{2014ApJ...794..132R,2017NatAs...1E.131R}. We showed that the magnetic modification of the planetary-scale equatorial waves causes the superposition of the magneto-Kelvin and $n=1$ magneto-Rossby waves to reverse in structure in the strong field limit. Hence we used arguments of simple linear wave dynamics to explain the magnetic wind reversal mechanism. 

Understanding the magnetic-reversal mechanism in terms of a shallow MHD phenomenon provides information about the magnetic fields on HJs. Repeating the numerical analysis of \sectref{sect:numerical:treatment:solutions} in the parameter spaces of HAT-P-7b and CoRoT-2b, we find that the minimum toroidal field strengths sufficient to magnetically reverse winds are $B_{\phi, \text{HAT-P-7b}} \gtrsim 3 \,(P / \onebar )^{1/2} \kilogauss $ and $B_{\phi, \text{CoRoT-2b}} \gtrsim 1 \, (P / \onebar)^{1/2} \kilogauss$,  where $P$ is the atmospheric pressure/depth of the reversal and the ideal gas law is used to convert from velocity units. These minima can be linked to dipolar field strengths using the scaling laws of \cite{2012ApJ...745..138M}, yielding $B_{\di{dip},\text{HAT-P-7b}}  \gtrsim 6 \gauss $  and $B_{\di{dip}, \text{CoRoT-2b}} \gtrsim 3 \kilogauss $. We comment that the striking difference between the two dipole field minima is a consequence of the temperature dependence of the magnetic Reynolds number \citep{2010ApJ...719.1421P,2012ApJ...745..138M}. The minimum dipole strength in the atmosphere of HAT-P-7-b agrees with the three-dimensional simulations of \cite{2017NatAs...1E.131R} and lies well below the range $50$--$100 \gauss$ predicted for most inflated HJs \citep{2017ApJ...849L..12Y}. The dipole field strength necessary to magnetically reverse the winds on CoRoT-2b ($3 \kilogauss$) greatly exceeds $250 \gauss$, the maximum surface dipole estimate for HJs \citep{2017ApJ...849L..12Y}. We conclude that wind reversals on HAT-P-7b are highly likely to be magnetically driven, whereas other explanations such as cloud asymmetries \citep{2013ApJ...776L..25D,2016A&A...594A..48L,2016ApJ...828...22P,2017ApJ...850...17R} or asynchronous rotation \citep{2014ApJ...790...79R} appear more plausible on CoRoT-2b.

There are several interesting questions that we do not address in this Letter. First, it is unclear how a highly temperature dependent (and hence horizontally varying) magnetic  Reynolds number will effect the toroidal-poloidal scaling relationship, and hence the dynamics of the wind reversal process. Furthermore, vertical magnetic fields have also been assumed to be small compared to horizontal fields in our model. Three-dimensional simulations are required to avoid this approximation.

\acknowledgments
\mygray{We acknowledge support from STFC for A.W.~Hindle's studentship, the Leverhulme grant RPG-2017-035 and thank Andrew Gilbert, Andrew Cummings, and Natalia G\'omez-P\'erez for useful conversations leading to the development of this manuscript.}


\end{document}